\documentstyle[prb,aps,preprint,epsfig]{revtex}

\begin{document}
\title{Nonequilibrium spectral diffusion due to laser heating 
in stimulated photon echo spectroscopy of low temperature glasses}


\author{Peter Neu$^{\ast}$, Robert J. Silbey$^{\ast}$, 
Stephan J. Zilker$^{\dagger}$, and Dietrich Haarer$^{\dagger}$\\
{\it $^{\ast}$Department of Chemistry and Center for Materials Science and Engineering,
 Massachusetts Institute of Technology,  Cambridge, Ma.  02139}\\
{\it $^{\dagger}$Physikalisches Institut und Bayreuther
Institut f\"ur Makromolek\"ulforschung, Universit\"at Bayreuth, D-95440 Bayreuth, Germany\\
{\footnotesize (Submitted to PRB)}}}

\maketitle
\thispagestyle{empty}

\begin{abstract} 
A quantitative theory is developed, which accounts for heating artifacts in three-pulse
photon echo (3PE) experiments. The heat diffusion equation is solved and the average value of 
the temperature in the focal volume of the laser is determined as a function of the 3PE waiting
time. This temperature is used in the
framework of nonequilibrium spectral diffusion theory
to calculate the effective homogeneous linewidth of an 
ensemble of probe molecules embedded in an amorphous host. The theory fits recently observed 
 plateaus and bumps   without  introducing 
a gap in the distribution function of flip rates of the two-level systems or any other
major modification of the standard tunneling model.
\end{abstract}


\section{Introduction}
Glasses are commonly considered as nonequilibrated, structurally disordered solids.
The amorphous network performs configurational fluctuations on a wide range of time scales.
At low temperatures, tunneling through conformational barriers is assumed to be the dominant
relaxation mechanism. This view is equivalent to the two-level system (TLS) model \cite{Phil,AHV}
 that has successfully
accounted for many of the anomalous low-temperature properties of 
glasses.\cite{P} In the standard
formulation, it is assumed that TLS have  on a logarithmic time scale
a broad distribution of energy splittings, $E$, and 
relaxation rates, $R$,
\begin{equation}\label{Dist1}
P(E,R)\, dE\, d\log R\  \sim \ {\rm constant}\,\times \,  dE\, d\log R.
\end{equation}
 As a consequence, the observed relaxation dynamics in glasses 
depends on the experimental time scale. This has far-reaching implications  for the linewidth
of an optical transition of a probe molecule  embedded in a glassy host: its transition frequency
becomes a dynamical quantity.\cite{FH} This phenomenon is commonly 
referred to as spectral diffusion (SD).
It prevents  a definition of a homogeneous linewidth  as in crystalline materials.

In an optical experiment, the absorption spectrum of an ensemble of chromophores doped into an
amorphous host is measured. Due to inhomogeneities in the local strain or electric fields
experienced by individual chromophores, this spectrum (the so-called inhomogeneous line) 
is several orders of magnitude broader than the linewidth of a single molecule. Hence, to monitor
the local structural glass dynamics, 
 line-narrowing techniques like
 two-, three-pulse (stimulated) photon echoes
(2(3)PE) and hole burning (HB)  have to be used.\cite{Volker,BF,LF,Haar2}
Being conceptionally similar, the main difference of these techniques is the sensitivity
to relaxation processes on different time scales: fast dephasing processes (pico--nanoseconds)
caused by rapid TLS flips and vibrations of individual 
chromophores\cite{JS} are measured in 2PE; 
linewidths measured in 3PE (nano--milliseconds) and HB (microseconds--several days)
 are additionally broadened due to SD. The rates of SD may vary over many orders of magnitude.
By convention, the linewidth measured in a 2PE experiment is called homogeneous. In the 
standard model, Eq. (\ref{Dist1}), a linear increase with temperature is predicted.
 The linewidth measured in 
3PE and HB experiments is commonly referred to as ``effective'' homogeneous linewidth. According
to Eq. (\ref{Dist1}), this line also 
broadens linearly with temperature. By the same token, SD induces
 a logarithmic broadening as a function of the 
 waiting time (time between the second and third pulse in a 3PE, or 
between burning and reading a hole in a HB experiment).\cite{HW,BH,Re,SS}

Very recently, there has been broad interest in 3PE of
Wiersma's group\cite{Wiers}$^{(a)-(e)}$, 
which---contrary to the standard model---showed a plateau 
between 10$^{-6}$ and  10$^{-3}$~s. They investigated the chromophore zincporphine in 
deuterated ethanol glass (ZnP/EtOD). Meijers and Wiersma
 interpreted their data by postulating 
a gap in the distribution of TLS rates in the
 microsecond region.\cite{Wiers}$^{(b)-(e)}$ Zimdars and Fayer
emphasized, shortly after that, the contradiction
 between hole burning data of Fayer's group\cite{LF}
and the 3PE data of Wiersma's group.\cite{Wiers}$^{(b)-(e)}$ 
To explain this difference, they invented
 a model, which is based on the assumptions that (1) the coupling between the chromophores
and the glass can vary as a function of the TLS relaxation rates, and (2)  TLS with
relaxation rates in the microsecond domain couple to the chromophores only via an electric
dipolar interaction. According to this,  non-polar chromophores such as those used in Wiersma's
experiment will not sense a perturbation by those electric TLS resulting in a gap  in the 
relaxation rate distribution. The relevance of this model has been discussed 
by Silbey and co-workers. \cite{SV}$^{(b)}$ The same authors could fit 
Meijers' data qualitatively
without assuming a gap, but starting out from a modified distribution function, which
predicts a slower-than-logarithmic line broadening in the
 microsecond region.\cite{SV}$^{(a)-(b)}$ 
Silbey's model is based on computer simulations of a Lennard-Jones model glass.\cite{HS}
Finally, it was Zilker and Haarer who shed light into the discussion by showing
in the first 3PE experiment {\it below} 1~K that the line broadening does not only level off in
the microsecond region, but in fact decreases again after several hundred microseconds.\cite{ZH}
They investigated the chromophore zinc meso-tetraphenylporphine in 
polymethylmethacrylate  (ZnTPP/PMMA), a polymer glass.
As the authors pointed out, this reversibility indicates laser induced sample heating artifacts,
which may also be responsible for the gap previously observed in Wiersma's group.

It is the content of this work to clarify this point by treating the laser induced
heating effect in a quantitative way, and reanalyzing the 3PE data of Meijers and Zilker.

The paper is organized as follows: in Sec. II, we brievely describe the 
experimental apparatus used in Ref.~\CITE{ZH} and give an estimation 
of the heating effects for  EtOD and PMMA  using the  details of 
Meijers' and Zilker's experiment;
in Sec. III, we present a simple model
for the time dependence of the temperature in the focal volume of the laser beam
by solving the heat diffusion equation for appropriate boundary conditions; in Sec. IV,
we calculate the optical linewidth in the framework of 
nonequilibrium spectral diffusion\cite{F1};
in Sec. V, we compare with Meijers' and Zilker's 3PE data, and, in Sec. VI, we conclude.

\section{Experimental}

\subsection{Stimulated photon echoes}
The stimulated photon echo measurements in Ref. \CITE{ZH} (see Fig.~1)
 were performed with a tunable dye laser
 (Coherent 702-1CD) synchronously pumped by a mode-locked argon-ion laser
 (Coherent Innova 200-10). The output pulses (6~ps width) are cavity-dumped
  and amplified in two dye cell amplifiers which use a two-stage design.
 Each chain is pumped by its own frequency-doubled nanosecond Nd:YAG laser
 (Spectron SL404G) at a repetition rate of 10~Hz. 
After each stage a saturable absorber is used to suppress amplified spontaneous
 emission. Two beams with equal intensities are obtained 
from the resulting 1~$\mu$J pulse of the first amplifier 
 by a beam splitter. Beam 2 is fixed in time, whereas beam~1
 can be delayed via a motorized delay line (DL1). The third pulse can be electronically 
delayed by picking any pulse from the 76~MHz pulse train emitted from the dye laser. 
Synchronization between the dye and the YAG lasers is accomplished via the reference 
signal of the mode-locker electronics, which triggers a digital delay generator 
(Stanford Research Systems DG535). The latter controls the firing of the flash 
lamps and the Q-switches of the YAG lasers and is also used to trigger the cavity-dumper 
of the dye laser. This design allows us to create pulses which can be electronically 
delayed with respect to each other from 100~ns up to 100~ms. An optical delay line 
(DL3) is used to realize waiting times shorter than 7~ns.  The optical density of 
the sample was about 1 at the wavelength of the study, 597~nm. The laser was 
focused on the sample to a spot size of approximately 100~$\mu$m; the pulse 
energy was about 3~nJ. The good optical quality of the samples enabled us to perform
 measurements of stimulated photon echoes without using an optical 
upconversion scheme. \cite{wiersma1:78}
The signal is spatially filtered to block most of the scattered light, and  
separated from the excitation beams by diaphragms. Finally, it is  
detected by a photomultiplier tube and registered with a boxcar 
integrator. The echo intensity depends nonlinearly on the energies 
of the excitation pulses. This makes an A/B--measurement
for compensation of excitation energy fluctuations impossible.
Therefore only data points, for which the energy of each laser
pulse---measured by the two photodiodes (D)---is within a $\pm$10~$\%$ window,
 are processed by the computer. Several scans of the delay line were
 averaged for each signal curve.

Zinc meso-tetraphenylpor\-phine (ZnTPP, Porphyrin Products Inc.) was dissolved at 
a concentration of $1.5\times 10^{-4}$~M in the monomer met\-hylmetha\-crylate; the
 polymerization of the mixture was initiated at 50$^{\circ}$~C. The resulting samples
 had a thickness of 1~mm and a radius of 4~mm.
 The fluorescence lifetime $T_1$ of ZnTPP in 
PMMA was measured as 7.5~ns in a transient-grating experiment. \cite{Z2}

The sample 
 was mounted in a $^3$He bath cryostat (Janis Research Co.) providing a
 temperature range from 0.4 up to 4.0~K.
The temperature was measured with a calibrated Germanium resistor
 (Lake Shore GR-200A-100) attached to the copper sample holder. Temperature
 control was achieved with an accuracy of $\pm$0.01~K by heating 
the internal charcoal pump of the cryostat.

\subsection{Heating effects in PMMA and EtOD}
Let us assume that at $t=0$ the total sample is in equilibrium with the
helium bath at a temperature $T_0$ and illuminated by the (first and the
second) laser beam. The laser irradiation excites via $S_0 \to S_1$ transitions
those chromophores of the sample which are in its  focal volume
$V_{\rm focal} = \pi a^2 L_p$. The latter is a cylinder given by the spot size
$a \approx  50\,\mu$m and the penetration depth  $L_p \approx  0.5$~mm of the laser beam.
 The sample is  cylindrically shaped with
 a radius $R_0$ of 4~mm and thickness  $L = 1$~mm. The chromophores are
assumed to be homogeneously distributed in the illuminated volume, and the laser
beam propagation is taken to be perpendicular to the circular surface of the sample.
The excited chromophores irradiate heat, $Q$,  over the fluorescence lifetime 
$\tau_{fl}\approx 1-10$~ns (in PMMA: $\tau_{fl}\approx 7.5$~ns, 
in EtOD: $\tau_{fl}\approx 2.7$~ns)
due to  intersystem crossings,
$S_1 \to T$, to the triplet state. Since metalporphines such as ZnP 
and ZnTPP undergo these transitions 
with probabilities of more than 90~$\%$ ISC yield, up to 25~$\%$ of the incident laser energy 
can be transformed into heat during this process.  If the heat release is considerably 
faster than the heat diffusion process out of the focal volume,
we can assume that after a time $\tau_{fl}$ the chromophores have
heated up the entire focal volume to a maximum temperature $T_1$ which is
determined by the relation
\begin{equation}\label{4}
Q = \int_{T_0}^{T_1} c(T) dT.
\end{equation}
In PMMA, the specific heat is given by\cite{SH} $c(T) =
 4.6\, T \,\mu$J/gK$^2$~+~$29.2\, T^3 \,\mu$J/gK$^4$.
Under the conditions of Zilker's 3PE experiment\cite{ZH} 
(pulse energy of 3 nJ, ISC yield of 97~\%, 
$\Delta E({\rm singulet-triplet}) = 3800$~cm$^{-1}$) 
for ZnTPP/PMMA,  $Q = 20\ \mu$J/g is 
released into the sample which results in a local temperature increase inside
the focal volume of $\Delta T = T_1 - T_0 \approx 0.55$~K for $T_0 = 0.75$~K.
At $T_0 = 1.5$~K the heating effects amounts into $\Delta T \sim 0.2$~K, and no 
heating effect is expected at 3~K.

In EtOD glass, the specific heat is not known below 2~K. The  value at two 2~K 
is roughly by a factor two larger than in PMMA: 
 $c_{\rm EtOD} (2\, {\rm K}) \approx 350$~$\mu$J/gK. \cite{VR}
The chromophore used in Meijers' experiment, ZnP, has very similar optical
properties as ZnTPP, which was used in Zilker's experiment:  ISC yield of 98~\%, 
$\Delta E({\rm singulet-triplet}) = 3700$~cm$^{-1}$.\cite{R} 
The applied pulse intensity in the 3PE experiment of Meijers and Wiersma has been 
 quoted to be ``less than 200~nJ per pulse'' in Ref. \CITE{Wiers}(a) and ``less than
100~nJ per pulse'' in Ref. \CITE{Wiers}(d). Our estimation of the heating temperature,
$T_1$,  is  based on a pulse intensity of 100~nJ/pulse.
 Since the optical 
properties of both chromophores are nearly
identical, one can estimate the heat release into the sample in Meijers' 
experiment by multiplying
Zilker's result by 100/3, which amounts in $Q \sim 650\ \mu$J/g. In PMMA, this provides
a temperature increase of $\Delta T \approx 1.6$, 0.8, and 0.6~K at $T_0 = 1.75$, 2.4, and 3~K,
respectively. Since the specific heat in EtOD is roughly by a factor 2 larger than in PMMA,
one has to divide these numbers by 2, yielding $T_1 \approx 2.55$, 2.8, and 3.3~K at 
$T_0 \approx 1.75$, 2.4, and 3~K, respectively.

\section{Heat diffusion}
Our goal is to determine the heat flow out of the focal volume
 by solving the heat diffusion equation 
\begin{equation}\label{D}
{\partial \over \partial t}  T({\bf r},t) -
 D \nabla^2 T({\bf r},t) = 0
\end{equation}
for the local 
temperature $T({\bf r},t)$, where the diffusion constant
\begin{equation}\label{3}
D = {\kappa\over \rho c}
\end{equation}
is controlled by the heat conductivity $\kappa$, the mass density $\rho$,
and the specific heat $c$ of the sample. To calculate
the line broadening, the spatially averaged temperature
$T(t) = \langle T({\bf r},t)\rangle_{{\bf r}\in V_{\rm focal}}$ will be, eventually,  plugged
in the formalism of nonequilibrium spectral diffusion. The
underlying assumption is  that the phonons are in quasi-thermal equilibrium
with the momentary temperature $T(t)$ in the focal volume, i.e., the TLS
relaxation rates $R(T)$ are given by $R[T(t)]$. This is reasonable, because the
phonon relaxation time is much shorter than the 
diffusion times (\ref{1})---(\ref{Kapitza}).  
The replacement $T(r,t) \to T(t)$ is clearly approximate. In a correct treatment,
the $r$ dependence of the temperature has to be handled on the same footing as the
$r$ dependence of the dipole-dipole interaction ($\sim r^{-3}$) between 
the chromophore and the TLS in the host, which would, however, render the model too complicated.

There are three time scales for the heat diffusion process:
the radial and longitudinal diffusion time
over the time scales
\begin{eqnarray}\label{1}
\tau_a &=& a^2/D,\\
\label{2}
\tau_L &=&  L_p^2/D,
\end{eqnarray}
 and the time scale over which heat
radiates  into the helium bath [note $c({\rm He})\gg c({\rm PMMA/EtOD})$]
\begin{equation}\label{Kapitza}
\tau_r = R_K \rho c V_{\rm focal},
\end{equation}
which is controlled by the Kapitza surface resistance between a solid and liquid helium
\begin{equation}
R_K = {0.05\over \pi a^2 T^3} \left[{{\rm K}^4 {\rm m}^2\over{\rm W}}\right].
\end{equation}
 Although $\kappa$ and $c$ depend
on temperature, we will assume in the following that $D$ is temperature
independent. Orders of magnitude for $\tau_a$, $\tau_L$, $\tau_r$
 are given for PMMA  in  Table I.

Stimulated echo experiments are performed 
on the time scale between 0.1--1~ns to 10--100~ms.
The plateau or the bump appears in the data between about
10~$\mu$s and  100~$\mu$s. This suggests the following picture: For $t < \tau_r \sim 2$~ms
heat cannot radiate into the helium bath due to the large value of the boundary 
resistance $R_K$. Accordingly, the probe itself operates  as a heat bath and 
absorbs the heat flow out of the focal volume in direction radial 
to the laser beam. At $\tau_L\gg\tau_a$ any inhomogeneous
 illumination  of the focal volume in longitudinal direction becomes unimportant. 
The radial flow sets the time scale
for equilibration with the helium bath after about $10 - 100\   \mu$s. At this time scale,
we would expect a bump in the stimulated echo data. 
After several hundred microseconds, we further
expect a return to   equilibrium spectral diffusion at $T = T_0$. Triplet heating
can be excluded,  since $\tau_r$ is much less than the time period between two
consecutive  echo pulse series ($\sim 75$~ms).
Both experiments
confirm this picture (cf. Fig. 2 and 3). 

Let us now perform our analysis more quantitatively. To solve the diffusion equation,
appropriate boundary conditions have to be posed. First, to keep the model simple
we neglect any heat conduction in propagation of the beam. This might be a crude 
approximation, however being justified by  $\tau_a \ll \tau_r$
and because the laser beam penetrates almost the whole sample (penetration length $L_p \sim 0.5$~mm).
The heat conduction in direction radial to the beam is governed by diffusion
into the sample. Hence, we have to solve the diffusion problem in an infinite 
(boundary effects due to a finite   $R_0 = 4$~mm are not important in this experiment)
cylindrical medium in which   heat is produced  by point sources (chromophores)
at a rate $\sim \Delta Q(r)  e^{-t/\tau_{fl}}$  ($r = \sqrt{x^2 + y^2}$).
  The amount of heat, $\Delta Q(r)$, released
by the chromophore at position $r$ is governed by the laser intensity, 
$I(r)$, at that point in the sample.
 We assume that
the intensity profile across the diameter of the laser beam is Gaussian,
\begin{equation}
I(r) = P{e^{-r^2/2a^2}\over 2\pi a^2},
\end{equation} 
where $P$ is the total amount of beam power (in W) used. 
The exact solution to this problem is very complicated and beyond the scope of this paper.
Instead we will treat a heat diffusion problem, which is equivalent in all physical aspects  to 
the problem mentioned above.
We assume that the chromophores have heated  the focal volume  to a higher 
temperature $T_1$  after 
 the time $\tau_{fl}$.  If $\tau_{fl} \ll \tau_a$, the temperature profile is then well approximated
by the laser intensity profile,
\begin{equation}\label{profile}
\Delta T(r,0) =  2 \Delta T \, e^{-r^2/2a^2}.
\end{equation}
The normalization, we have chosen, is such that the average temperature in the focal volume at $t = 0$
is $\Delta T$, i.e., $\Delta T = \int_0^\infty dr\, 
2\pi r \Delta T(r,0)\, e^{-r^2/2a^2}/2\pi a^2$.
The temperature
at a later time, $t$, is given by integration over the heat kernel\cite{CJ}
%
%
\begin{eqnarray}\label{Sr0}
\Delta T(r, t)  =  {1\over 4\pi Dt}\int_{-\infty}^{\infty} dx' \int_{-\infty}^{\infty} dy'\,
\Delta T(x',y', 0)\, e^{-[(x-x')^2 + (y - y')^2]/4Dt}  = 2\Delta T 
 {e^{-r^2/2[\sigma(t)]^2}\over  [\sigma (t)]^2/a^2},
\end{eqnarray}
%
%
where
\begin{equation}
\sigma(t) = \sqrt{2D t + a^2}.
\end{equation}
Hence the average temperature 
$\Delta T(t) = \int_0^\infty dr\, 2\pi r \Delta T(r,t)\, e^{-r^2/2a^2}/2\pi a^2$ 
at a later time, $t$, reads
\begin{equation}\label{Sr}
\Delta T(t) = {\Delta T\over 1 + t/\tau_a}.
\end{equation}
Eventually, to account for the delay of the heat release over the time scale 
$\tau_{fl}$, we subtract $e^{-t/\tau_{fl}}$ from Eq. (\ref{Sr}). This  corresponds to assuming
that heat is produced at the rate $(Q\rho/\tau_{fl}) e^{-t/\tau_{fl}}$ 
per unit time and unit volume
 yielding a final temperature increase\cite{CJ}
  $\Delta T = Q/c$.
This, finally, gives
\begin{eqnarray}\label{Tt}
T(t) = T_0 + (T_1 - T_0) \,\left\{ {1\over 1 + t/\tau_a} \ - \ e^{-t/\tau_{fl}}\right\}.
\end{eqnarray}
Of course, this equation is only an approximation for the real
 solution of the diffusion problem posed above.
As a consequence, the numerical value for $\Delta T$ and $\tau_a$ obtained from the fits
by using (\ref{Tt}) have to be seen as an order of magnitude estimation.
However, our model contains all physical characteristics: (i) heat is supplied 
over the time scale $\tau_{fl}$ with
an exponential rate, (ii) the temperature returns 
algebraically  (non-exponentially)  to equilibrium after the time scale $\tau_a$.
We have checked that (i) and (ii)  are independent of the initial 
temperature profile (\ref{profile}).
For instance $\Delta T(r,0) \propto \Theta(a - r)$ gives the same kind of behavior.

\section{Optical linewidth}
Hu and Walker \cite{HW} and Suarez and Silbey \cite{SS} have shown that three-pulse
photon echo amplitude decays as $\exp[-\langle F_1(\tau) + F_2(\tau,t)\rangle_{\rm TLS}]$
due to TLS-flipping  in the amorphous host. Here
$\langle F_1(\tau)\rangle_{\rm TLS}$ is the 
two-pulse echo decay (or dephasing decay), and  $\langle F_2(\tau,t)\rangle_{\rm TLS}$ 
is the decay that depends on the separation, $t$,  between the second and
the third pulse. $\langle ...\rangle_{\rm TLS}$ denotes the ensemble average over
the TLS-relaxation rates, $R$, and the TLS-energy splittings, $E$.
If $\tau$ lies well within a hyperbolic distribution of relaxation rates $R$,
 i.e., $R_{\rm min} \ll 1/\tau \ll R_{\rm max}$ and $P(R) \sim 1/R$, the
effective, i.e., $t$ depending  dephasing rate $[1/T_2](t)$ is defined by 
$\langle F_1(\tau) + F_2(\tau,t)\rangle_{\rm TLS} = 2\tau/T_2(t)$. Analogously, the 
two-pulse echo decay rate, $1/T_{2,{\rm 2PE}}$  is defined by
$\langle F_1(\tau)\rangle_{\rm TLS} = 2\tau/T_{2,{\rm 2PE}}$. 
Subtracting the lifetime contribution defines the pure dephasing rate 
$1/T^\ast_{2,{\rm 2PE}} = 1/T_{2,{\rm 2PE}} - 1/2T_1$.
The effective homogeneous  linewidth measured in a 3PE experiment is defined by
\begin{equation}
[1/\pi T_2^\ast](t) =   
[1/\pi T_2^*]_{\rm 2PE} + 
[1/\pi T_2^\ast]_{\rm SD}(t)
\end{equation}
where the last term, arising  from $F_2(\tau,t)$,
 is the  contribution due to spectral diffusion.
For TLS and phonons in thermal equilibrium at temperature $T_0$ with 
 a hyperbolic distribution in TLS-relaxation rates, $R$,
 and a flat distribution in the TLS-energy splitting, $E$, (cf. Eq. (\ref{Dist1}))
 one finds the well-known result 
\begin{equation}\label{eqSTM}
[1/\pi T_2^\ast](t) =   K T_0\, [\, 3.66 \  +  \ \log (R_{\rm eff} t)\,]\ + \ \Gamma_{\rm PLM}.
\end{equation}
Here,  $K$ is a
collection of constants proportional to the dipole-dipole interaction strength
between the TLS and the chromophore, and $R_{\rm eff}$ is an effective relaxation rate
averaged over the TLS splittings, $E$, (see below). 
The second term, $[1/\pi T_2^\ast]_{\rm SD}(t) \propto \log(R_{\rm eff} t)$, 
is the waiting time dependent contribution    due to spectral diffusion. 
The first and the last term 
describe pure dephasing. At very low temperatures the 
term linear in $T$ is dominant. This term arises from  $F_1(\tau)$ and is characteristic for
TLS-induced dephasing in amorphous solids.  The last term, characterized 
 by an activated temperature dependence
\begin{equation}\label{PLM}
\Gamma_{\rm PLM} =  b\, {\exp(-\Delta E/k_BT)\over [1 - \exp(-\Delta E/k_BT)]^2},
\end{equation}
also appears in molecular mixed crystals. It is a pure dephasing mechanism 
arising from   vibrational modes of individual chromophores (pseudo-local phonon modes).\cite{JS}
This mechanism usually starts to be dominant in the temperature regime  between 1 and 3~K.

Recently, Silbey et al.\cite{SV}$^{(a)}$ have  introduced a 
new model based on molecular dynamics simulations on a Lennard-Jones computer glass.\cite{HS}
In the simulations for Ni-P at a given composition,
 it was found that the distribution function $P(\Delta)$ 
of the tunneling frequency, $\Delta$, is hyperbolic for $\Delta/k_B$ less than $\sim 10^{-3}$~K,
but as $\Delta$ increases, $P(\Delta)$ is best fitted by a form $1/\Delta^{1-\nu}$ where $\nu$
increases to $\sim 0.2$ as $\Delta/k_B$ increases beyond 1~K.
 With a typical maximum relaxation rate
of about $10^{10}$~s$^{-1}$ at 1~K, this implies that the standard model ($\nu = 0$)
does not apply for waiting times shorter than 
$\sim 10^{-3}-10^{-4}$~s. Although these results are not
quantitatively accurate for all glasses, we will see that they provide a  more reliable picture
in the investigation of heating effects in the 3PE data below. 
The distribution function of TLS-relaxation rates and energy splittings is modified in this model
according to 
\begin{equation}\label{SModel}
P(E,R) = {\rm constant} \times  
{E^\nu (R/R_{\rm max})^{\nu/2} \over R  \sqrt{1-R/R_{\rm max}}}.
\end{equation}
(We set Silbey's second parameter $\mu\equiv 0$.\cite{SV}$^{(a)}$)
This yields a two-pulse decay varying as $\exp[-(2\tau/T_{2,{\rm 2PE}})^{1-\nu/2}]$, which is 
very close to an exponential decay for fits of experimental data. In the limit $\nu \ll 1$, 
the effective  homogeneous linewidth  is then 
found to behave like
\begin{eqnarray}\label{eqSM}
[1/\pi T_2^\ast](t) &=& 
    K_\nu  T_0^{1+\nu} \left(\, \Theta_\nu  \  +\  (2/\nu) [1 - (R_{\rm eff} t)^{-\nu/2}]\right)\nonumber\\
&+& \Gamma_{\rm PLM},
\end{eqnarray}
where 
\begin{equation}
R_{\rm eff}^{-\nu/2} = 
\frac{\int_0^{E_{\rm max}} dE\, E^\nu \,{\rm sech}^2(E/2k_B T)\,R_{\rm max}^{-\nu/2}}
{\int_0^{E_{\rm max}} dE\, E^\nu \,{\rm sech}^2(E/2k_B T)},
\end{equation}
and (replacing $E_{\rm max}/2k_BT$ by infinity)
\begin{equation}
K_\nu =  K\, 
\int_0^\infty dx\,  x^\nu {\rm sech}^2(x)
\end{equation}
with $\Theta_{\nu = 0} = 3.66$ and 
$\Theta_{\nu = 0.25}\approx  3.4$  (note $K_{\nu = 0} = K$).

To calculate the spectral diffusion contribution in the case of heating over the
time scale $t$, it is more convenient to use the formalism of Black and Halperin.\cite{BH}
According to this,  the contribution to
the optical linewidth due to spectral diffusion is determined by the number
of TLS, $n_f(t)$, which have flipped an odd number of times during  the time
$t$. Denoting the initial occupation of the upper (lower) state by $n_+(0)$
$(n_-(0))$, and the upwards (downwards) relaxation rate by $W_+$ ($W_-$),
a simple master equation approach yields
\begin{equation}\label{8}
n_f(t) = \frac{n_-(0)W_+ + n_+(0)W_- }{R} (1 - e^{-t R}),
\end{equation}
where $R = W_+ + W_-$. In our case, all TLS are initially in thermal
equilibrium with the helium heat bath at $T_0$. Hence,
$n_+(0) = [\exp(E/k_BT_0)+1]^{-1}$ and $n_-(0) = [1 + \exp(-E/k_BT_0)]^{-1}$.
If we assume quasi-thermal equilibrium for the phonons, i.e., if we assume
that the phonons are at the momentary temperature (\ref{Tt}), the relaxation
rates become time dependent and satisfy
$W_+(t)/R(t) =  [\exp(E/k_BT(t))+1]^{-1}$ and
$W_-(t)/R(t) = [1 + \exp(-E/k_BT(t))]^{-1}$, with the one-phonon rate
\begin{eqnarray}\label{9}
R[r,T(t),x] &=& r R_{\rm max}[T(t),x], \\
R_{\rm max}[T(t),x]  &=& c\, T_0^3 x^3 \coth[x T_0/T(t)],
\end{eqnarray}
where $x = E/2k_B T_0$ and $c \approx 10^{10}$~K$^{-3}$~s$^{-1}$.
In the experimentally relevant parameter regime, the exact numerical solution
of the master equation with time dependent rates is very exactly described by
the following analytical expression:
%
%
\begin{eqnarray}\label{10}
n_f(t) =  (1/2) \{ 1 - \tanh[E/2k_BT_0] \tanh[E/k_BT(t)]\}\,
\left(1-e^{- t R[r,T(t),x]}\right).
\end{eqnarray}
The average of  $n_f(t)$ over the TLS-energy splitting,
$E$, and the dimensionless relaxation rate, $r\equiv R/R_{\rm max}$, is now 
performed using the distribution function (\ref{SModel}).
With this, the effective homogeneous linewidth at the momentary temperature $T(t)$ reads
\begin{eqnarray} \label{11SM}
[1/\pi T_2^\ast](t) &=&   
 K_\nu  T_0^{1+\nu} \left[ \, \Theta_\nu  +     f_\nu(t, R_{\rm max}[T(t)])\, \right] 
 \, + \,   \Gamma_{\rm PLM},
\end{eqnarray}
where
\begin{eqnarray}
 f_\nu(t, R_{\rm max}[T(t)])  = 
{\int_0^{\infty}  dx\, [1 - (t\, R_{\rm max}[T(t),x])^{-\nu/2}]\,
[1 - \tanh(x) \tanh(x T_0/T(t))] \over 
(\nu/2)\int_0^{\infty}  dx\ x^\nu\,  {\rm sech}^2(x)}.
\end{eqnarray} 
%
%
The result for the standard model arises in the limit $\nu \to 0$.
 In Fig. 2 and 3, $[1/\pi T_2^\ast](t)$ [(\ref{eqSM}) and (\ref{11SM})]   is shown
together with 3PE data of Ref. \CITE{Wiers}(d)  for ZnP in EtOD,
and  Ref. \CITE{ZH}  for  ZnTPP in PMMA. In Fig. 4, 2PE data
of ZnTPP  and  tetra-tert-butyl-terrylene  (TBT) in PMMA are shown together with fits made with
\begin{equation}\label{2PE}
[1/\pi T_{2}^\ast]_{\rm 2PE} =  K_\nu\,  T_0^{1+\nu}  \Theta_\nu  +  \Gamma_{\rm PLM}.
\end{equation}

\section{Comparison to  experiment}
Meijers and Wiersma\cite{Wiers} have studied the 3PE decay of various chromophores 
in a number of different glasses. In order to fit their data they were forced to modify the
standard model (\ref{Dist1}) by postulating the existence of a gap in the distribution
of TLS flip rates on the time scale between micro- and  milliseconds.
For the system ZnP/EtOD, they assumed that  the decay rate is linear in $\log (t)$ from the
earliest times ($\sim 10^{-10}$~s)  to $10^{-6}$~s, flat until $10^{-3}$~s, and then once more
linear  above  $10^{-3}$~s, with the same slope as in the early time regime. As discussed in 
Ref. \CITE{SV}(a),  the physical nature of this gap remains totally unclear. 

Zilker and Haarer\cite{ZH} have studied for the first time a 3PE decay below 1~K. They found
at 0.75~K for the system ZnTPP/PMMA that the decay rate increases from the
earliest times ($\sim 10^{-10}$~s)  to $10^{-6}$~s, levels off, and {\it
decreases} after  $\sim 300\ \mu$s. As the authors noted, this reversibility clearly
indicates laser heating artifacts in the sample. Clearly, the question arises
whether laser heating might also turn out to be responsible
for the observation of a plateau in Meijers'  and Wiersma's data.

In Ref. \CITE{SV}(a)-(b),
Meijers' and Wiersma's data have been analyzed with the modified standard model using
Eq. (\ref{eqSM}). The authors have found qualitative  agreement with better quality fits
emerging  for the high temperature data. Zilker and Haarer\cite{ZH} also analyzed 
successfully their data at 1.5 and 3~K, but not at 0.75~K, within this model.

In Fig. 2(a), we show the 3PE data for ZnP/EtOD at three temperatures, taken from the thesis of 
Meijers,\cite{Wiers}$^{(d)}$ with two fits using Eq. (\ref{eqSM}), dashed dotted line,  and 
 Eq. (\ref{11SM}), full line. As noted already in 
Ref. \CITE{SV}, the need for a gap in the distribution
function has disappeared. As shown in the figure, the predicted value of the 2PE decay rates 
also agree with experiment.
One clearly sees from this plot, that including heating artifacts
into the theoretical description improves the fits to the 1.75 and 2.4~K 
data, significantly. We have 
demonstrated this more clearly in Fig. 2(b).   
The initial temperature increase due to laser heating is $\Delta T = 0.9$, 0.4, and 0.3~K at
$T_0 = 1.75$, 2.4, and 3~K, respectively. Both 3PE and 2PE show consistently that 
the contribution to pure dephasing of pseudo-localized librational modes of 
the guest chromophores in the
amorphous matrix is  insignificant even at  3~K for ZnP in  EtOD glass.\cite{Wiers}$^{(c)}$

In Fig. 3(a), the 3PE data for ZnTPP/PMMA taken from Zilker and Haarer\cite{ZH} are plotted along
with our theory in the same way. Again the data can be fitted with no gaps in the distribution 
function, and with parameter values close to those used in Figs. 2(a)-(b). 
The reversibility of
the line broadening for the 0.75~K data is well reproduced by our theory, 
as depicted in Fig. 3(b).
This  clearly shows the influence of laser heating artifacts.
No heating effect is observed at 3~K data; the temperature increase  at 0.75 and 1.5~K is
found to be 0.75 and 0.2~K, respectively.  However, the predicted value of the 
2PE decay rates is much lower than the experimental value. 
The 3PE data  for ZnTPP/PMMA can  well 
be fitted without (or with only a small, cf.  Fig. 4)  
 contribution from localized phonon modes.
 Contrary to this, the 2PE data 
clearly show a deviation from a power law temperature dependence above $
\sim 2$~K, which indicates
active localized vibrations of the chromophores.
In Fig. 4, we have illustrated this using 
recently published 2PE data for ZnTPP/PMMA \cite{Z2} together with new results for TBT/PMMA.
The full line  shows
a fit of Eq. (\ref{2PE}) with  $\nu = 0.18$. The best fit is 
obtained with $\nu = 0$ (dashed line).
The dashed dotted line  demonstrates the discrepancy between  the experimental data and 
 Eq. (\ref{2PE}) when the parameter set obtained from the 3PE data (Fig. 3) is used.
 This may indicate that there are extra dephasing processes in 
these systems other than those described 
by the present model.

The parameters for the heating process are summarized in  Table II.
There  is close agreement between the estimations  for $T_1$ 
presented in Sec. II and the fit values. For PMMA, 
the fit value of the diffusion time $\tau_a$ is 
about one order of magnitude  larger than  the estimations in Table I.
 For pulse intensities  less than 100~nJ in Meijers' and Wiersma's experiment
the agreement between estimated (Eq.~\ref{4}) and fitted  value for $T_1$
is less close, although the discrepancies are insignificant given the uncertainties
in the experimental parameters for EtOD.

It must be stressed that a fit to the 3PE data
is also possible, if one uses  the standard model with $\nu = 0$. 
However, the heating temperature
we had to assume  in this case were  too high. Since $\nu \neq 0$ introduces already 
a weak bending of the decay rates, it also brings down the value for $T_1$ to realistic values.
We consider this observation as an indication that it is necessary to change the standard model
in the short time regime. For longer waiting times, $\stackrel{>}{\sim} 10^{-3}$~s, the 
experiment explores that part of the distribution function, which becomes more and more
hyperbolic. Hence, the bending of the decay rates should become weaker and, eventually,
merge into the  linear growth with $\log (t)$ usually observed in HB experiments. We believe
that the deviation of the data from our theory in the 
millisecond regime indicates this transition.

\section{Conclusions} 
We have extended Silbey's modified standard model for dephasing and spectral diffusion 
in optical experiments to the situation of a nonequilibrated phonon bath. In particular,
we have studied the situation in which the  focal volume of a laser beam in an amorphous host
is heated to some temperature $T_1$, and cools down to the temperature of the helium bath
by heat diffusion. The resulting nonlogarithmic line broadening of chromophore molecules
embedded in the glass due to spectral diffusion has been calculated. Our motivation has been 
to investigate whether this picture explains recently observed plateaus and bumps 
in 3PE data. We conclude that laser heating effects are important for 3PE experiments
in the kelvin regime. This together with a modified, slightly bended equilibrium line 
broadening quantitatively accounts for the experimental data. However, since no studies 
of the 3PE decay on laser fluence are available up to now, the experimental evidence in favor
of heating artifacts is indirect. A study of the fluence dependence remains an {\it essential}
 experiment to test the significance of heating effects.

\acknowledgements 
This research has been supported   by the National Science Foundation, the 
Alexander von Humboldt Foundation, and  the 
Deutsche Forschungsgemeinschaft, SFB279. We also thank  Daan Thorn Lesson,
David R. Reichman. Frank L. H. Brown,  and Yu. G. Vainer
for discussions.




%
%


\newpage
\section*{Figure Captions}
\begin{itemize}
 \item[FIG. 1:] Experimental setup for  stimulated photon echo measurements. 
                 ``50/50'' marks 50~$\%$~beam splitters, DL1--3 are optical delay lines, 
                 D photodiodes, and  PMT is a photomultiplier tube.

\item[FIG. 2:](a) 3PE decay rates, $[1/\pi T_2^*](t)$, 
               as a function of the waiting time, $t$, for 
                 ZnP  in EtOD,  from Ref. \CITE{Wiers}. The lines are fits 
                   to  the data:
                   (---) heating effects included  [Eq. (\ref{11SM})],
                ($- \cdot -$) no heating effects included 
                    [Eq. (\ref{eqSM})].  Effective relaxation rates used are 
               $R_{\rm eff} =  1.2 \times 10^{9}$s$^{-1}$ ($\circ$),
               $R_{\rm eff} = 2.8 \times 10^{9}$s$^{-1}$ ($\ast$),
               $R_{\rm eff} = 3.5 \times 10^{9}$s$^{-1}$ ($\times$). 
             The parameters for the heating process  are $\tau_a = 15\ \mu$s,
               $T_1 = 2.65$~K  ($\circ$), $T_1 =  2.8$~K $(\ast)$,  
                and  $T_1 = 3.3$~K $(\times)$.
               The values of the 2PE decay rates are also given 
                in the figure ($+$) along with the prediction of the model (dotted lines)
                with  $K_\nu = 26.9$~MHz/K$^{1+\nu}$.
               The contribution of $\Gamma_{\rm PLM}$ is neglected in the 
		fits at all three temperatures.
               (b) Same as (a) but only the 
               data at $T_0 = 1.75$~K are shown.
               
 \item[FIG. 3:] (a) 3PE decay rates, $[1/\pi T_2^*](t)$, as a function of 
		the waiting time, $t$,  
               for  ZnTPP  in PMMA,  from Ref. \CITE{ZH}.  The lines are fits
                   to  the data:
                   (---) heating effects included  [Eq. (\ref{11SM})],
                ($- \cdot -$) no heating effects included 
                    [Eq. (\ref{eqSM})]. Effective relaxation rates used are 
               $R_{\rm eff} =   10^{11}$s$^{-1}$ ($\circ$),
               $R_{\rm eff} = 1.3 \times 10^{11}$s$^{-1}$ ($\ast$),
               $R_{\rm eff} = 2.6 \times 10^{11}$s$^{-1}$ ($\times$). 
             The parameters for the heating process  are $\tau_a = 300\ \mu$s,
               $T_1 = 1.5$~K  ($\circ$), $T_1 =  1.7$~K $(\ast)$,  
                and  $T_1 = T_0$ $(\times)$.
               The values of the 2PE decay rates are also given 
                in the figure ($+$) along with the prediction of the model (dotted lines)
                with  $K_\nu = 15.2$~MHz/K$^{1+\nu}$, $b = 0.9$~GHz, and
                $\Delta E = 9.7$~cm$^{-1}$. (b) Same as (a) but only the 
               data at $T_0 = 0.75$~K are shown.
              
  \item[FIG. 4:] 2PE decay rate, $[1/\pi T_2^*]_{\rm 2PE}$,
               for ZnTPP $(\circ)$ and TBT $(\ast)$
               in PMMA. The solid line is a fit of   Eq. (\ref{2PE}) to the data for
               $\nu = 0.18$, $K_\nu = 24$~MHz/K$^{1+\nu}$, $b = 15.8$~GHz, and
                $\Delta E = 9.7$~cm$^{-1}$. The dashed line is  a fit of   Eq. (\ref{2PE})
               with $\nu = 0$, $K = 25.7$~MHz/K, $b = 12.7$~GHz, and
                $\Delta E = 8.8$~cm$^{-1}$. The dashed dotted line shows Eq. (\ref{2PE})
               with  parameter values obtained from the fit to  the 3PE data [Fig. 2(a)]:  
              $\nu = 0.18$, $K_\nu = 15.2$~MHz/K$^{1+\nu}$, $b = 0.9$~GHz, and
                $\Delta E = 9.7$~cm$^{-1}$.           
\end{itemize}

\section*{Table Captions}
\begin{itemize}
\item[Table I:] Diffusion times in PMMA with $\rho = 1.18$~g/cm$^3$, $c =
 4.6\, T \,\mu$J/gK$^2$~+~$29.2\, T^3 \,\mu$J/gK$^4$ (Ref. \CITE{SH}), and 
$\kappa = (15, 20, 30, 42, 50) \times 10^{-3}$~W/mK 
(Ref. \CITE{Nittke}) at different temperatures $T_0$.

\item[Table II:]  Heating parameters.

\end{itemize}

\section*{Tables}

Table I:
$$
\begin{tabular}{l|lll}
 \hline\hline
$T_0$ [K] &$ \quad \tau_a$  [$\mu$s] &$\quad \tau_L$ [ms] & 
$\quad \tau_r$  [ms]   \\ \hline\hline
0.75    & $ \quad$  3.10   &$ \quad$  0.15   & $ \quad$  2.21          \\
1       & $ \quad$ 4.99    &$ \quad$  0.25   & $ \quad$  1.99              \\
1.5     & $ \quad$ 10.37   &$ \quad$  0.53  & $ \quad$  1.84          \\
2       & $ \quad$ 17.05   &$ \quad$  0.85  & $ \quad$  1.79             \\
3       & $ \quad$ 43.03   &$ \quad$  2.15  & $ \quad$  1.75       \\ \hline\hline     
\end{tabular}
$$

Table II:
$$
\begin{tabular}{l|llll}
 \hline\hline
     & $T_0$ [K] $\quad$ & $T_1^{\rm est}$ [K] $\quad$ & $T_1^{\rm fit}$ [K]  $\quad$& $\tau_a$ $[\mu$s]\\   \hline\hline
PMMA  & 0.75      & 1.3                 & 1.5         & 300       \\
     & 1.5       & 1.7                 & 1.7          &      \\
     & 3         & 3                   & 3            &      \\\hline
EtOD &  1.75     & 2.55                & 2.65         & 15\\
     &  2.4      & 2.8                 & 2.8          & \\
     &  3        & 3.3                 & 3.3         &\\ \hline\hline
\end{tabular}
$$

\newpage

\section*{Figures}

FIG. 1.

\begin{figure}[hbt]
\epsfxsize 12 cm
\centerline{\epsffile{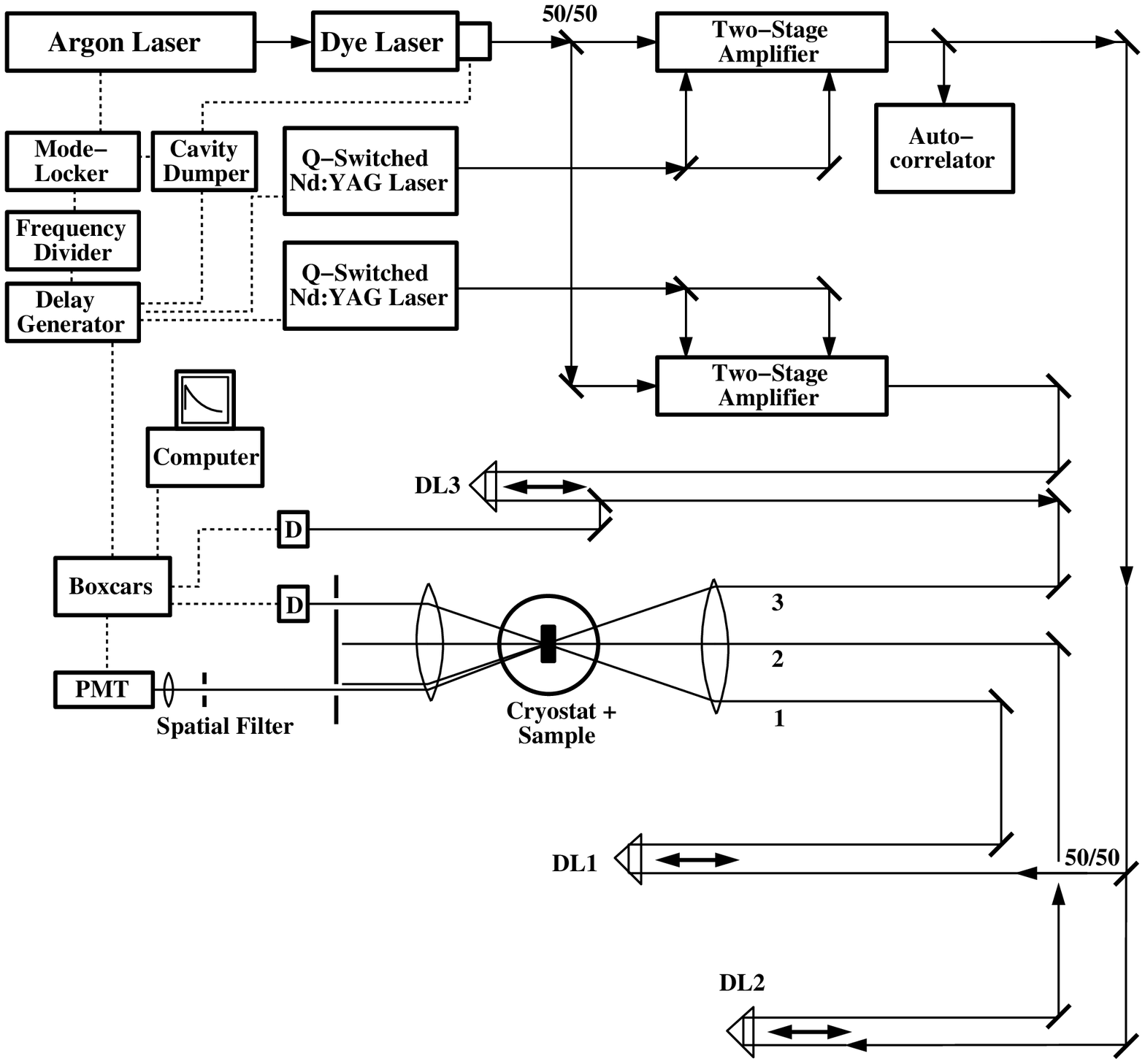}}

\end{figure}

\newpage

FIG. 2.
\begin{center}
               \epsfig{file=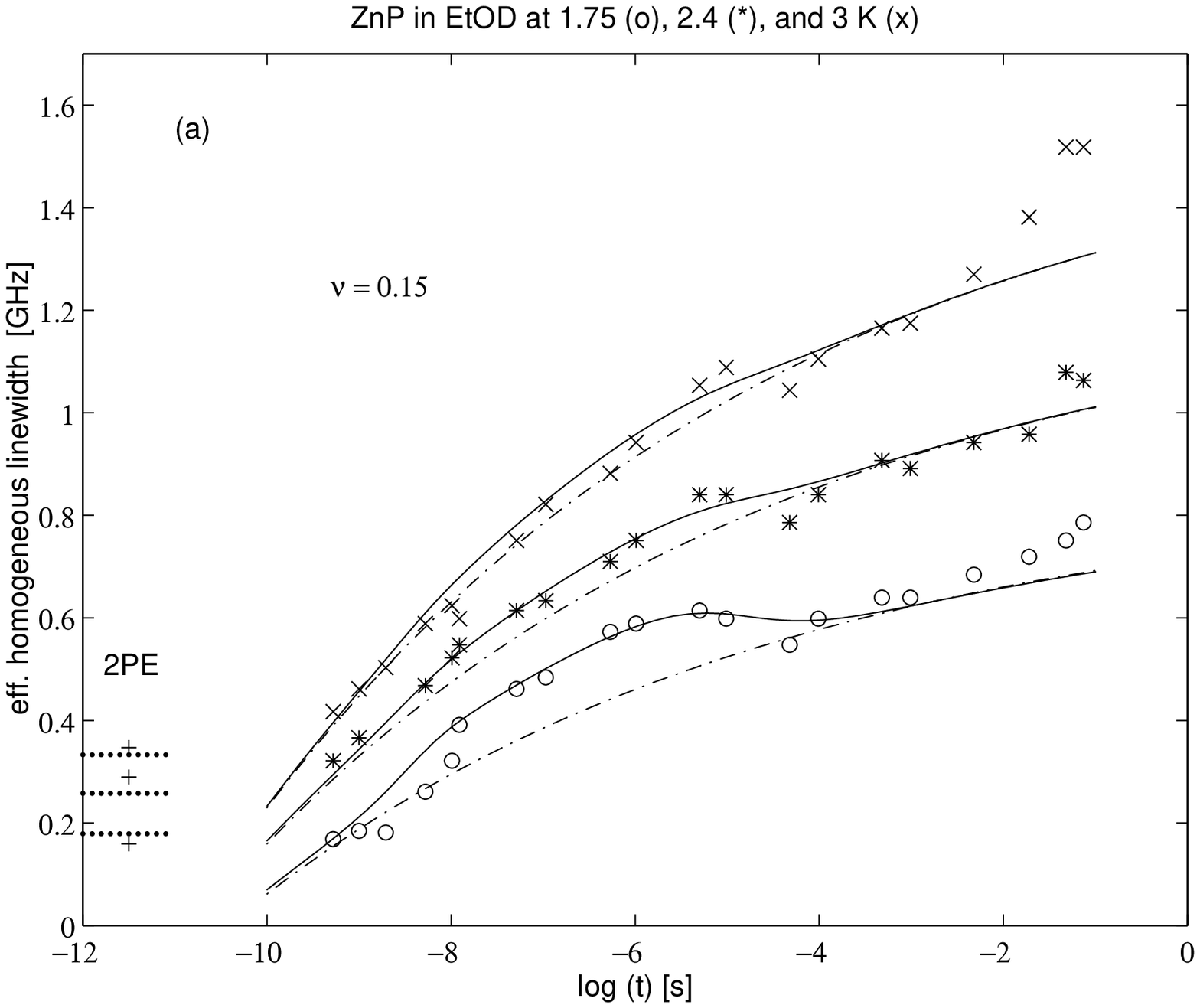}
                \end{center}
               \begin{center}
               \epsfig{file=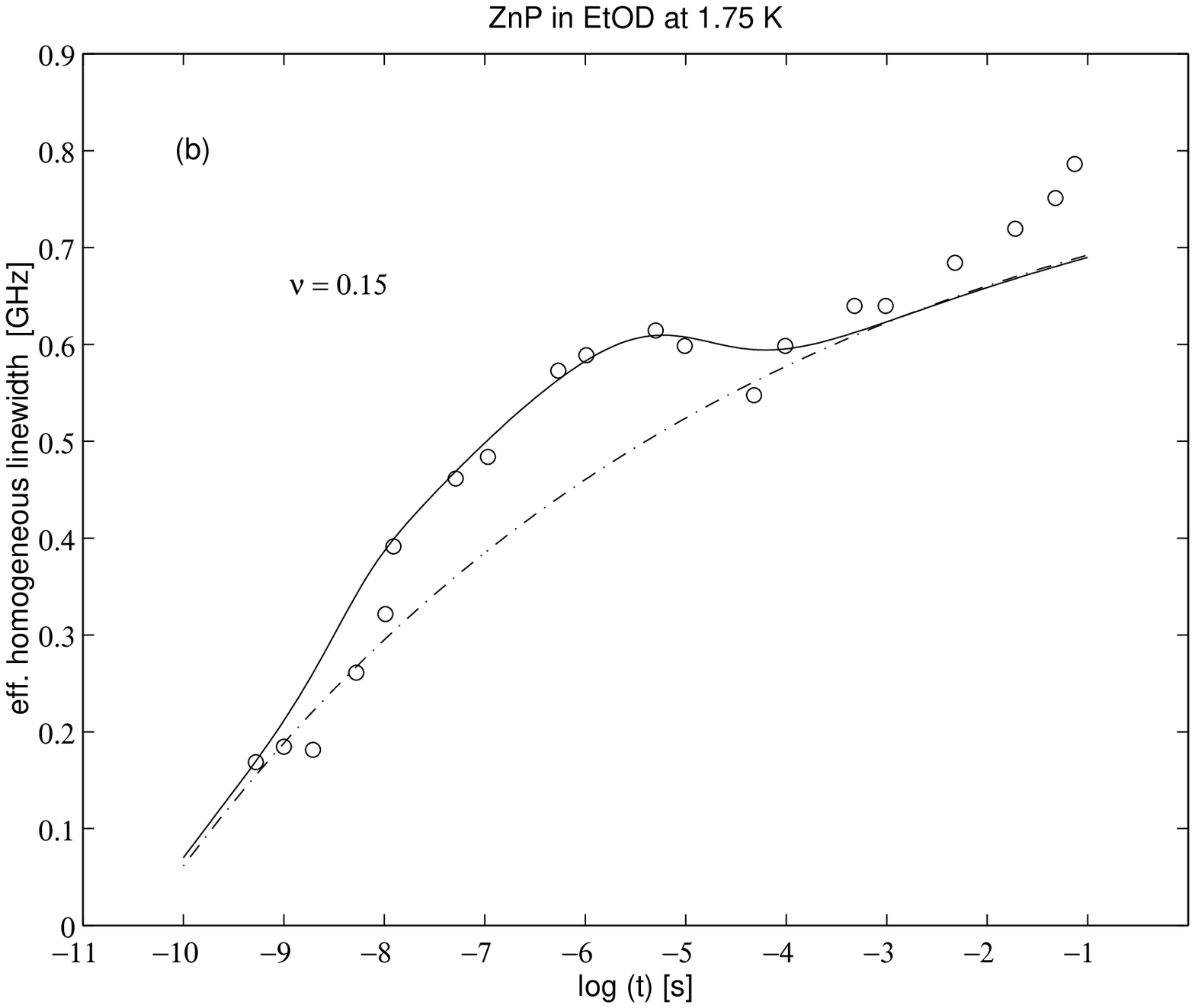}
\end{center}

\newpage

FIG. 3.
 \begin{center}
               \epsfig{file=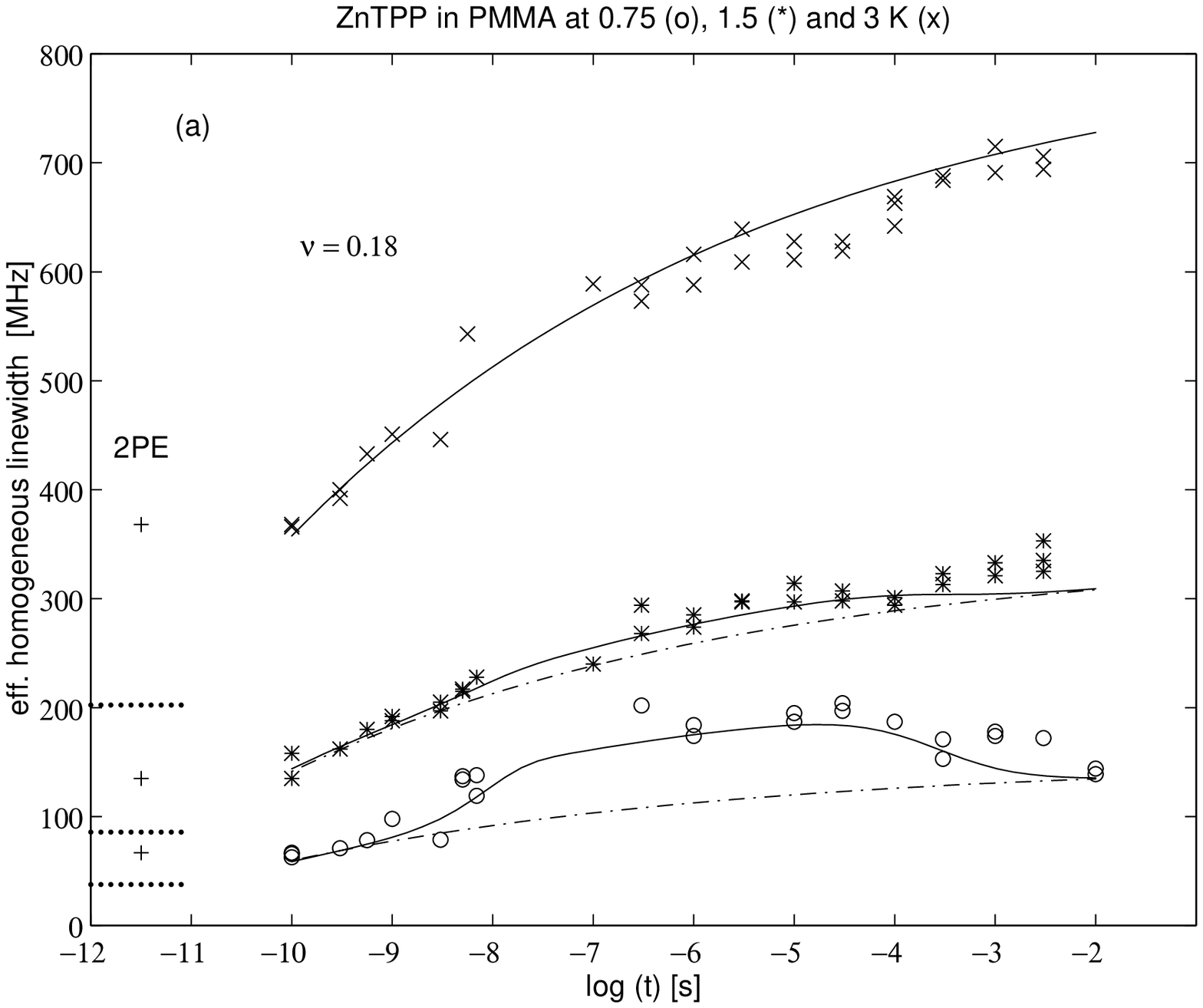}
                \end{center}
                 \begin{center}
               \epsfig{file=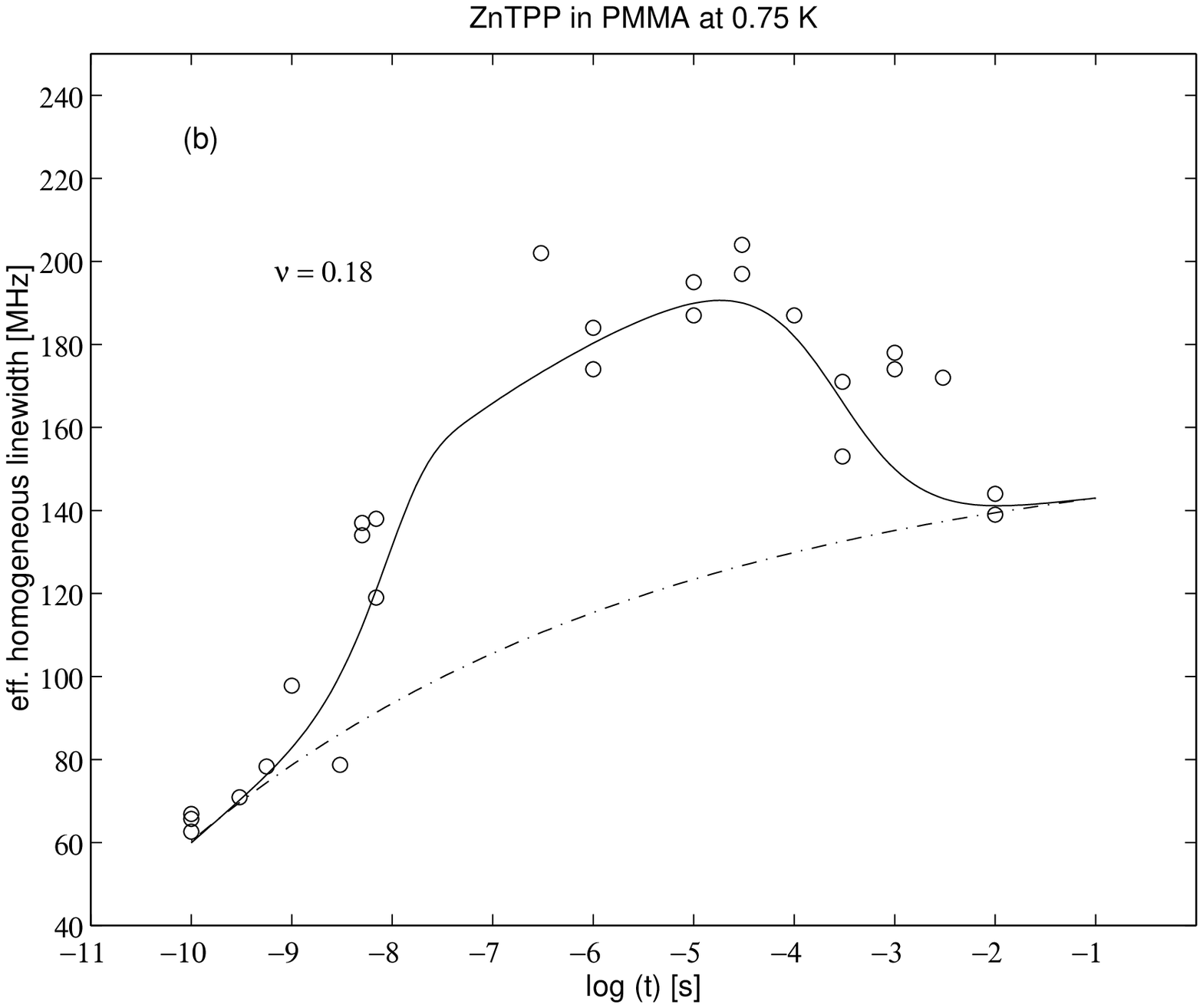}
\end{center}

 \newpage  

FIG. 4.                              
              \begin{center}
               \epsfig{file=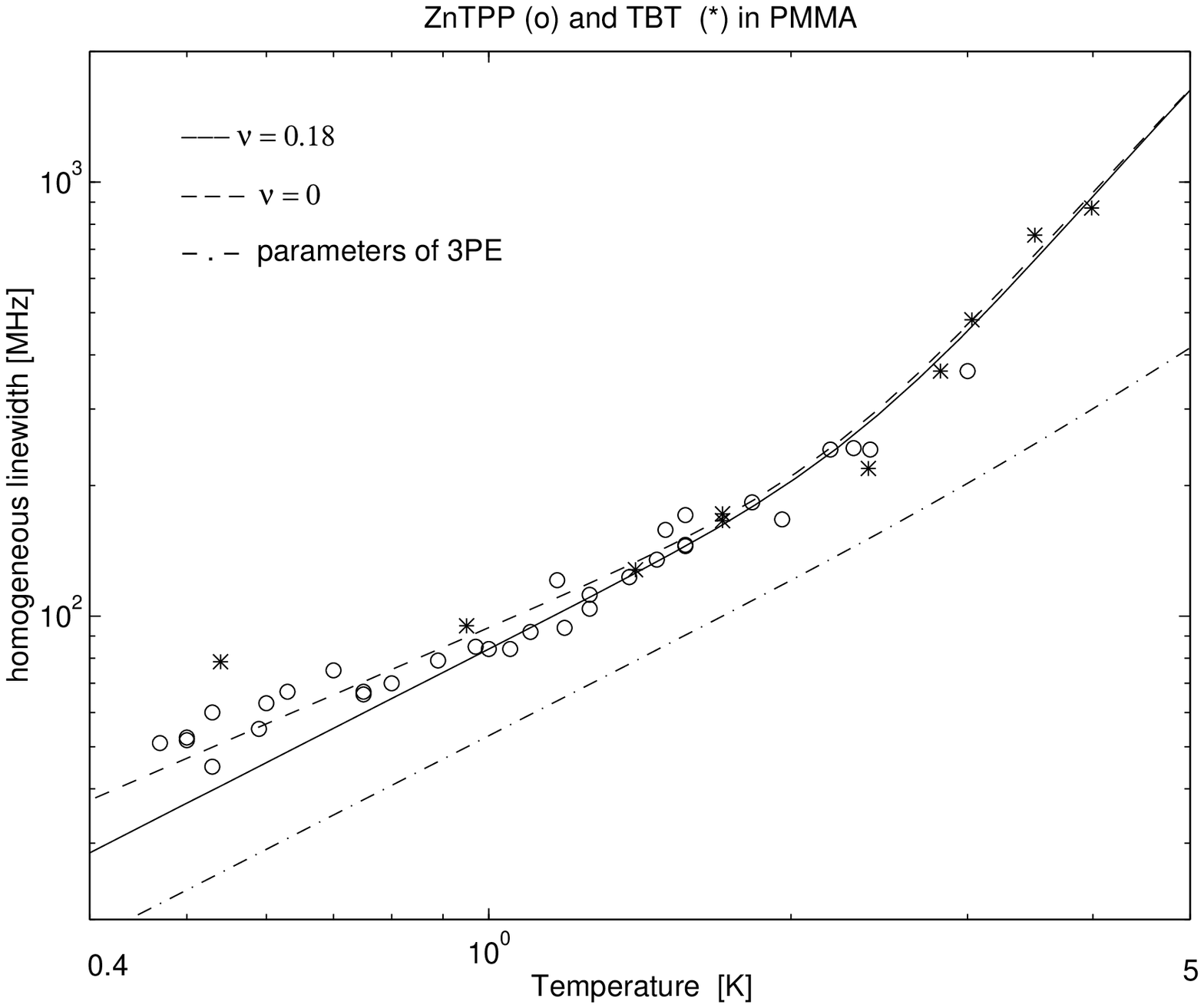}
                \end{center}
\end{document}